\documentstyle[psfig,conf_iap,10pt]{article}
\begin{document}
\heading{%
%
The IGM at high redshift and galaxy formation. \\
%
} 
\par\medskip\noindent
\author{%
Alain Blanchard$^{1}$, Simon Prunet$^{2}$
}
\address{%
Observatoire astronomique de Strasbourg, ULP, 11, rue de l'universit\'
e,
67 000 Strasbourg,  France
}
\address{%
Universit\'{e} de Paris-Sud,
Institut  d'Astrophysique Spatiale, B\^{a}timent 121,\\
F-91405 Orsay Cedex, France}
%
\begin{abstract}
The conditions for structure formation which ultimately lead to galaxies
request further ingredients behind the simple collapse criteria. The Jean's 
criteria and the cooling criteria are those which are currently used. 
However in such a simple scheme, a fundamental problem occurs in 
hierarchical pictures, namely the 
{\sf overcooling}: the predicted fraction of primordial gas expected to 
have cooled 
in the history of structure formation   is for too large. The solution to 
this problem is likely to be a substantial re-heating phase. 
Here, we discussed one possible solution: the warm IGM picture. If 
the feedback of galaxy formation is able to heat the IGM up to temperatures
 of the order of $10^5-10^6$ K, galaxy formation is inhibited on small 
 mass scale. This leads to an inverse hierarchical picture: most of the large 
 galaxies form at  redshifts in the range  3 to 5, while small galaxies form
 at two different epoch: at an early phase at redshift greater than five and 
 at a late phase, between redshift 3 and 0. Such a scheme may reproduce quite 
 well the amount of HI gas versus redshift.
\end{abstract}
\section{Introduction}
The problem of galaxy formation is a central problem of cosmology. 
Recent pogresses, both from the theoretical side and the observational side,
have triggered numerous works. The most dramatic  changes are probably 
the ability to have access to direct information at high redshift: 
the HI gas, the star formation rate, the possible detection of an 
infrared background originating from early galaxies \cite{pug} as well as the 
direct spectroscopy of high redshift field galaxies provide a number of 
observational constraints to which theories can now be confronted.
The model we present  has a amazing small number of 
free parameters, and still reproduces quite well several key 
observational constraints.
\section{Recipes for galaxy formation}
\subsection{The global picture}
It is generally believed that structure formation originated accordingly 
to the 
gravitational instability picture. Structures which achieved a high contrast 
density, namely greater than 200, are called virialized.
Properties of clusters of galaxies can actually be used as 
useful  constraints on cosmological scenarios \cite{Jim}. The dynamics of 
dark matter 
seems to be understood well enough that the basic time evolution of the 
correlation 
function and the mass function of cosmic structures can be described at any 
redshift provide  that the power spectrum of the primordial fluctuations is 
known (\cite{ps}, \cite{ham}).
First attempts to address the question of galaxy formation
has met a first important apparent success:
it has been suggested that a criteria to differentiate 
dark halos leading to galaxies 
from those leading to clusters 
is the cooling criteria. When gas falls in a potential it is shock-heated
(and/or by adiabatic compression) up to the virial 
temperature allowing the gas
 to be in hydrostatic equilibrium.  Numerical simulations has confirmed 
\cite{gus}
that this simple argument provides an accurate estimation of the actual 
temperature:
\begin{equation}
  T_{v} = 5. 10^5 M_{12}^{2/3} (1+z) {\rm K} \label{eq:Tv}
\end{equation}
where $M_{12}$ is the total mass in unit of $10^{12} {\rm M}_{\odot}$
 of the object forming at redshift $z$. 
The typical size of the halo is:
\begin{equation}
  R_{v} = \left(\frac{T_{v}}{10^5 {\rm K}}\right)^{1/2}
\frac{1.}{(1.+z)^{3/2}} 45 h^{-1}{\rm kpc}
\label{eq:Rv}
\end{equation}
When the gas reaches its virial temperature,  
it has a characteristic cooling time.
Clusters typically represent 
structures for which the cooling time exceeds the age of the universe, while 
for galaxies it is much shorter. It is tempting 
therefore to conclude that the cooling criteria can be used as a criteria for 
star formation: if the gas is able to cool, it will contract in a runaway 
fashion, which is can end up only by star formation (as there is not so 
much cooled gas in the universe). This argument successfully explains the 
order of magnitude of the luminosity of the brightest galaxies ($L_*$).
\subsection{The overcooling problem}
\begin{figure}
\vspace*{-0.4cm}
\centerline{\hspace*{-1.5cm}
\vbox{
\psfig{figure=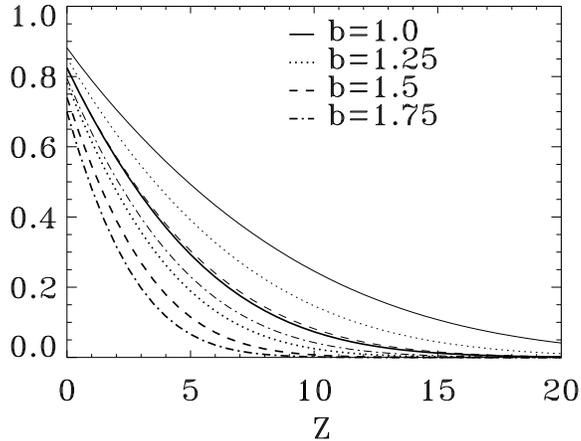,height=7.cm}
}}
\caption[]{Integrated fraction of gas able to cool at various redshift.
The different curves correspond to 
different values of the bias parameter 
($b=1.,1.25,1.75,2.$). The thin lines are for 
for the standard CDM ($\Gamma = 0.5$) 
while the  thick lines are for $\Gamma = 0.5$. \label{fig:Ftot}
}
\end{figure}

The previous scheme has remained a qualtitative picture for a while. 
However,  the need for a more quantitative picture has 
become clear as the amount of data on distant galaxies has 
increased: from the number counts of faint galaxies to the recent star 
formation rate versus redshift.
The first basic difficulty one faces on in the   
 simple cooling scheme is the so-called {\sf overcooling  problem}: 
at high redshift 
a large fraction  of the baryons lies is small potentials with temperatures
  in 
the range $10^{4}-10^{6}$K in which cooling is extremely efficient. 
Consequently, 
most of the baryons are expected to have been cooled by now. This is in clear 
contradiction with two basic facts : known stars 
represent only a small fraction of baryons predicted by nucleosynthesis,
typically 10\% 
and most of the baryonic content of clusters is still in the gas phase, while 
most of them should have been cooled.
Both facts suggest that only 10\% to 20\% of the 
primordial baryons were actually turn into stars 
during the cosmic history. This problem was first 
pointed out by Blanchard et al. \cite{bla} and Cole \cite{cole}. 
A simple estimate of the integrated cooled 
 fraction  can be obtained from the mass function of cosmic structures, 
 by noticing that any piece of gas 
 within a halo with $T_v > 10^4$K should have settled 
 in the cooling region at some earlier epoch  \cite{bla}: 
 \begin{equation}
  F_c(z) = \frac{1}{\rho}
\int_{m_4(z)}^{+\infty}N(m) m dm
\label{eq:Ftot}
\end{equation}
where ${m_4(z)}$ represents the mass of halos which have a virial 
temperature of $10^4$K at redshift $z$. The amount of total cooled gas at 
different redshift is presented in figure \ref{fig:Ftot}. The reality 
of this overcooling problem is not easy to test by mean of numerical 
simulations because of resolution limitations.
However, Navarro \& Steinmetz found that \ref{eq:Ftot} provide a reasonable
approximation and  therefore qiute reasonable to believe 
that the simple based Press and Schechter argument can be used.
Therefore, the  solution of the ovecooling problem 
implies that the gas have undergone some 
substantial reheating. The fact that the x-ray luminosity of clusters 
does not scale as predicted by the scaling argument provide a further 
evidence of a complicated baryon history. 
\subsection{The reheating phase}
The existence of a reheating phase of baryons has been advocated
in various contexts in galaxy formation scenario. For instance, White 
and Frenk \cite{wf} argued that the energy input of supernova from the 
first generation of stars is able to prevent the cooling of the gas 
that remains confined in galactic scale halos.  Blanchard et al.
\cite{bla} suggested a rather different picture :  the first objects
 which form heat the IGM to a 
temperature high enough that most of the gas does not fall in most of 
the forming potentials in which cooling would have been possible otherwise
because of the temperature of the gas. This introduces the 
idea that a key physical  quantity controlling 
 galaxy formation is the temperature 
of the IGM, which could be regulated by galaxy formation.
\subsubsection{A self-regulated IGM\\}
The basic equation which governs the temperature of the IGM in 
a self-regulated picture is:
 \begin{equation}
 (1-F_c)  \rho_b\Lambda(T(z))= \epsilon\dot{E_{*}}\frac{1}{\rho}
\int_{m_T(z)}^{+\infty}N(m) m dm
\label{eq:evolT}
\end{equation}
where  $T(z)$ is the temperature of the IGM at redshift $z$, $m_T(z)$ is
 the mass associated to this temperature, $ \dot{E_*}$ is the total 
 energy ouput resulting from star formation (essentially Supernova) 
and $\epsilon$  is  the energy fraction which is transferred to the IGM 
 and $\Lambda$ is 
 the cooling function. As the IGM is likely to be photo-ionized  at 
 the same time that it  undergoes the reheating we used the cooling 
 function of a photoionized gas. The temperature of the IGM versus redshift
 depends on the value of the efficiency 
 of energy injection to the IGM. The temperature of the IGM 
 is increasing from $10^4$K to reach 
 a maximum value between  $10^5$K and $10^6$K at redshift of the order of 
3 to 5 
 depending on the details of the model. Such a high temperature will easily 
 explain the absence of detected Gunn-Peterson decreement, even if most of 
 the baryons lies in the IGM. Explaining the existence of Lyman-$\alpha$ 
 clouds will be certainly challenging for this scenario: they could not 
 be small halos nor large scale fluctuations in the IGM. A possible 
 explanation might be that they are the extended parts of galactic disks. 
\section{From baryons to stars}
\begin{figure}
\vspace*{-0.4cm}
\centerline{\hspace*{-1.5cm}\vbox{
\psfig{figure=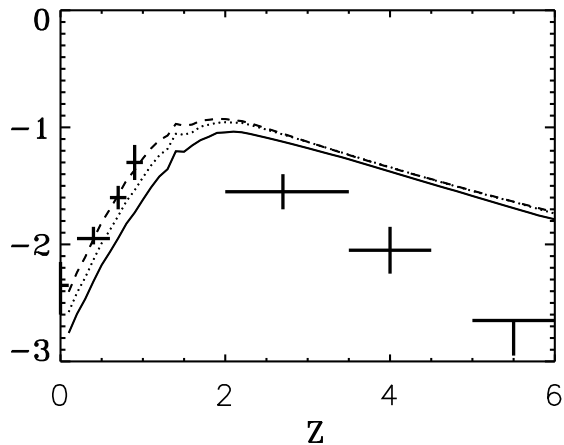,height=7.cm}
}}
\caption[]{ Star formation rate versus redshift in the warm 
photo-ionized picture assuming a delay of 2 Gyr.
The observational points are from \cite{mad}.\label{fig:mad}
}
\end{figure}

Although this model is relatively simple, there 
are still a number of free parameters. The power spectrum as well as the 
primodial nucleosynthesis value have been left free until now. The final 
important free parameter is the value of $\epsilon$. 
$\epsilon = 1.$ means that 
the energy transfer is 100\% efficient for a standard IMF 
(a higher value could 
be used, due to a non standard IMF or because of extra energy input). 
The parameter  $\epsilon$ can be constrained by computing the integrated 
amount of stars produced.  
\begin{figure}
\vspace*{-0.4cm}
\centerline{\hspace*{-.7cm}\vbox{ 
\psfig{figure=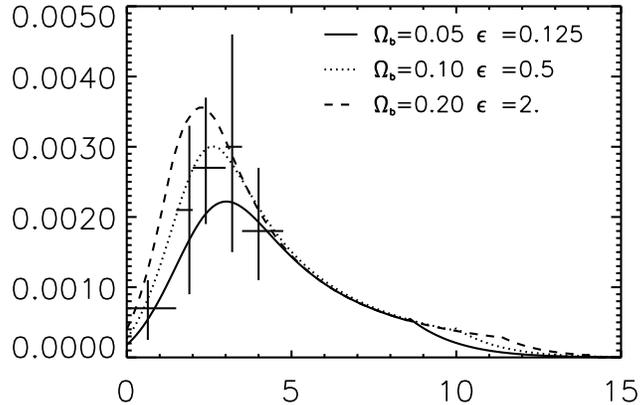,height=7.cm}
}}
\vspace*{-0.4cm}
\caption[]{ Theoretical amount of HI gas predicted in the  self-regulated  
photionized picture. 
The different curves correspond to 
different values of the  parameter $\epsilon$
($\epsilon= .125, 0.5, 2.$) used for different $\Omega_b$ 
($\Omega_b. = 0.05, 0.10, 0.20$) 
for the CDM-like spectrum with  $\Gamma = 0.25$.) 
\label{fig:HI}
}
\end{figure}
 The  gas which has been cooled can be compared directly 
 the amount of observed HI  gas versus redshift: this is presented 
 on figure \ref{fig:HI}. The models which are presented are those 
 who explained 
 the present amount of stars. At this stage, there are no  free parameters  
 other than the power spectrum. The standard CDM spectrum do not lead to
 the right  amount of HI versus redshift, while 
 the $\Gamma = 0.25$ fits impressively well the observed distribution.
 Given the relatively small number of free parameters, this is rather amazing.
 A further step can be obtained by noticing that the HI gas is likely to
 be the progenitors of present day stars (or at least of the progenitors 
 of stars in disks). Assuming that te HI gas is transformed  in stars but 
 with some delay, we can infer the star formation rate at different redshift 
 and compare it to the one inferred by Madau \cite{mad} from the CFRS 
 survey and HDF. This is illustrated by  figure \ref{fig:mad}. Such 
 modeling is rather crude, but is still rather instructive: we found that the 
 rapid decrease of star formation rate between redshift 1 and 0 is 
 expected in the self-regulated photoionized picture. Moreover the amplitude 
 can  be well reproduced, provide that star 
 formation from cooled HI is delayed by 2 Gyr. 
 The high redshift star formation rate is not well reproduced: the theoretical
  model systematically predicted a higher star formation rate. It is 
  fundamental to realize that this is due to the fact the Madau star 
  formation rate integrated from redshift 5 to 0 
  cannot explain the total amount of present day stars, and therefore 
  either this formation rate has been underestimated or there is an other 
  period of earlier star formation at high redshift (which is not 
  expected in any of the models presented here).\vspace*{-3mm}
\section{Conclusion}
One of the strongest problem in the galaxy formation history is the so-called 
overcooling problem. It is likely that its solution and consequently 
the galaxy formation history is connected to the 
thermal (and chemical) history of the IGM. We have presented a simple 
global coherent picture of the stars formation history based on the hypothesis 
of a self-regulating mechanism. This simple model impressively succeeds in
explaining the whole set of present day observational constraints one can  
set  on galaxy formation theory. It is therefore interesting to 
investigate such a model in more details. 

\acknowledgements{We would like to thank the organizers of the 1997 IAP meeting for this exciting meeting, for the wonderful conference dinner and for their patience...\vspace*{-3mm}}
\vfill
\end{document}